\documentclass[fleqn,usenatbib]{mnras}

\usepackage{newtxtext,newtxmath}

\usepackage[T1]{fontenc}
\usepackage{ae,aecompl}

\usepackage{graphicx}	
\usepackage{amsmath}	
\usepackage{amssymb}	
\usepackage{xargs} 

\newcommand{\hmol}{H$_2$} 

\title[Infrared molecular hydrogen lines in  GRB hosts]{Infrared molecular hydrogen lines in  GRB host galaxies}
\author[K. Wiersema et al.]{\parbox{\textwidth}{
K. Wiersema$^{1,2}$ \thanks{E-mail: K.Wiersema@warwick.ac.uk},  
A. Togi$^{3,4}$, 
D. Watson$^{5}$, 
L. Christensen$^{6}$, 
J. P. U. Fynbo$^{5}$,
B. P. Gompertz$^{2}$,
A. B. Higgins$^{1}$,
A. J. Levan$^{2}$, 
S. R. Oates$^{2}$,
S. Schulze$^{7}$,
J. D. T. Smith$^{8}$, 
E. R. Stanway$^{2}$, 
R. L. C. Starling$^{1}$, 
D. Steeghs$^{2}$,
N. R. Tanvir$^{1}$
\vspace{0.4cm}}\\
$^{1}$ Department of Physics and Astronomy, University of Leicester, University Road, Leicester LE1 7RH, UK\\
$^{2}$ Department of Physics, University of Warwick, Coventry CV4 7AL, UK\\
$^{3}$ Department of Physics and Astronomy, The University of Texas at San Antonio, San Antonio, 1-UTSA Circle, TX 78249, USA\\
$^{4}$ Department of Physics and Earth Sciences, St. Mary's University, One Camino Santa Maria, San Antonio, Texas 78228, USA\\
$^{5}$ The Cosmic Dawn Center, Niels Bohr Institute, University of Copenhagen, Juliane Maries Vej 30, 2100 Copenhagen \O, Denmark \\
$^{6}$ Dark Cosmology Centre, Niels Bohr Institute, University of Copenhagen, Juliane Maries Vej 30, 2100 Copenhagen \O, Denmark \\
$^{7}$ Department of Particle Physics and Astrophysics, Weizmann Institute of Science, Rehovot 7610001, Israel\\
$^{8}$ Ritter Astrophysical Research Center, University of Toledo, 2825 West Bancroft Street, M. S. 113, Toledo, OH 43606, USA}

\date{Accepted XXX. Received YYY; in original form ZZZ}

\pubyear{2018}

\begin{document}

\label{firstpage}
\pagerange{\pageref{firstpage}--\pageref{lastpage}}
\maketitle

\begin{abstract}
Molecular species, most frequently H$_2$, are present in a small, but growing, number of gamma-ray burst (GRB) afterglow spectra at redshifts $z\sim2-3$, detected through their rest-frame UV absorption lines.
In rare cases, lines of vibrationally excited states of H$_2$ can be detected in the same spectra. 
The connection between afterglow line-of-sight  absorption properties of molecular (and atomic) gas, and the observed behaviour in  emission of similar sources at low redshift, is an important test of the suitability of GRB afterglows as general probes of conditions in star formation regions at high redshift. Recently, emission lines of carbon monoxide have been detected in a small sample of GRB host galaxies, at sub-mm wavelengths, but no searches for H$_2$ in emission have been reported yet. 
In this paper we perform an exploratory search for rest-frame $K$ band rotation-vibrational transitions of H$_2$ in emission, observable only in the lowest redshift GRB hosts ($z\lesssim0.22$). Searching the data of four host galaxies, we detect a single significant rotation-vibrational H$_2$ line candidate,  in the host of GRB\,031203. Re-analysis of {\em Spitzer} mid-infrared spectra of the same GRB host gives a single low significance rotational line candidate. The (limits on) line flux ratios are consistent with those of blue compact dwarf galaxies in the literature. 
New instrumentation, in particular on the {\em JWST} and the {\em ELT}, can facilitate a major increase in our understanding of the H$_2$ properties of nearby GRB hosts, and the relation to H$_2$ absorption in GRBs at higher redshift.
\end{abstract}




\section{Introduction \label{sec:intro}}

Gamma-ray burst (GRB) afterglow spectroscopy has shown great promise as a probe of gas and dust properties within, and near,  
star forming regions in distant galaxies (see e.g. \citealt{Schady} for a review). The bright afterglows serve as backlights with a simple (sometimes reddened) synchrotron spectrum, against which
atomic and molecular absorption lines are easily distinguished. In addition, the ultraviolet radiation of the rapidly fading afterglow excites meta-stable and fine structure atomic states. The variability of absorption lines from these transitions allows precise distance determination of the absorbing gas with respect to the GRB in cases with high signal-to-noise spectra (e.g., \citealt{Vreeswijk}). 
Long gamma-ray bursts (broadly speaking those with a duration longer than $\sim2$ s) are associated with the deaths of massive stars, and their rate is therefore thought to trace cosmic star formation, 
likely moderated by a metallicity threshold for formation (e.g. \citealt{PerleyShoals2} and references therein). This means that GRBs and their bright afterglows may form a valuable tool to locate, and study, actively star forming galaxies across cosmic time, unbiased by galaxy luminosity (except perhaps through a luminosity-metallicity dependence).
A key advantage offered by GRBs over quasars as backlights, is that the host galaxies can be studied in emission once the afterglows have faded,
through spectral energy distributions (to obtain stellar population parameters, e.g.~\citealt{PerleyShoals2}) and emission lines (to obtain element abundances and star formation rates, e.g.~\citealt{KruehlerXSHemission}).
These studies in emission complement the line of sight studies of afterglows, connecting the afterglow sight lines through their hosts with host galaxy integrated emission properties. Sample sizes of hosts and afterglow spectra are growing, and statistical studies of  metal abundances, stellar populations and dust properties are now possible, placing GRB host galaxies in the context of wider galaxy surveys (e.g. \citealt{Vergani}). An important component of the picture, though, the molecular content, is still poorly understood. 
Of particular interest is the \hmol\ molecule, which plays a key role in the processes of star formation. 

An additional advantage of host studies is that it is not limited to the subset of GRBs for which the optical afterglow is detected (provided the  host can be reliably identified, see e.g. \citealt{Perley020819B}). There is evidence that most of the GRB sightlines that pass by significant column densities of molecules also will contain large dust column densities and hence that such sightlines are underrepresented in the subset of GRBs with well-detected optical afterglows 
(\citealt{Prochaska080607h2}; \citealt{Kruehlerh2grb120815}). The peculiar case of GRB140506, for which  
CH+ molecules were detected in absorption along with very steep UV extinction further supports this point (\citealt{GRB140506}; \citealt{Heintz17}).

The homonuclear \hmol\ molecule does not have a dipole moment, so electric dipole transitions between levels with different vibrational quantum number ($\nu$) or rotational quantum number ($J$) in the ground state are forbidden. Quadrupole transitions, however, are allowed, and the pure rotational lines (with $\nu = 0-0$) are located at (mid-)infrared wavelengths, which makes them challenging to study, particularly for faint sources like GRB hosts. Of more interest to us are the rotation-vibration (hereafter ro-vibration) transitions in the ground state. In the following we adopt the standard notation, where the difference in $J$ is given by a letter (O, Q, S for $\Delta J = +2,0,-2$, respectively), followed by the final state $J$, and preceded by the vibrational transition (so 1--0 S(1) is $\nu=1-0, J=3-1$). The main ro-vibrational lines are located at near-infrared wavelengths. For example, the 1--0 S(0), 1--0 S(1) and 1--0 S(3) transitions  are located at 2.22, 2.12 and 1.96 $\mu$m, respectively, in the restframe. At low redshifts, these transitions can therefore be detected by ground-based observatories.

The first allowed transitions out of the \hmol\ ground state to an excited state are the Lyman and Werner bands, 
which require ultraviolet (UV) photons. These transitions have indeed been observed in a handful of GRB afterglow spectra (\citealt{Prochaska080607h2}; \citealt{Kruehlerh2grb120815}; \citealt{DElia120327molecules}; \citealt{Friis}, and possibly in \citealt{GRB060206}), where the redshifts of the GRBs shift these transitions from the UV to the optical domain. Identification and analysis is difficult: these transitions are located among the atomic hydrogen lines of the dense Lyman forest, and as such a fairly high signal-to-noise and spectral resolution is required to separate them. In very rare cases, highly diagnostic absorption lines of vibrationally excited \hmol\ are found 
(\citealt{Scheffer2009}; \citealt{Kruehlerh2grb120815}), that are located redwards of the Ly$\alpha$ line. The detection of Lyman-Werner lines, combined with fits to the atomic hydrogen Ly$\alpha$ line (in GRB sight lines often found as a strong, highly damped line, a damped Lyman absorber [DLA], 
\citealt{Jakobsson06}) has allowed estimates of the molecular gas fraction (integrated over the line of sight), and helps to place the relatively small GRB \hmol\ absorption sample in the context of the much larger sample of quasar DLAs (e.g., \citealt{Noterdaeme08}; \citealt{Noterdaeme}). 

The sightline selection function of long GRBs is arguably quite different from those of QSO DLAs (e.g. \citealt{ProchaskaDLAs}; \citealt{Fynbo08}; \citealt{Fynbolowressample}), which makes the long GRB afterglow H$_2$ sample especially valuable as a probe of high-redshift star forming regions. Of particular interest is that long GRBs trace cosmic star formation (e.g. \citealt{Greiner}), and therefore the \hmol\ absorption seen in afterglows may probe the conditions in star forming regions within (dwarf) star-forming galaxies at the peak of cosmic star formation ($z\sim2-4$). 

However, whilst afterglow sightlines likely probe regions near long GRBs in high mass star forming regions, which should be rich in \hmol\ (e.g. \citealt{Tumlinsonnondetections}), the low detection rate, the excitation state of the detected \hmol, and occasionally the association of the \hmol\ absorber with excited atomic metal fine structure lines, have shown that in several cases the \hmol\ absorbers
are likely located far from the star forming region in which the GRB progenitor resided (e.g. \citealt{DElia120327molecules}). 
The low detection rate of \hmol\ in afterglow spectra is puzzling. Several explanations have been put forward, that likely all play a role: dissociation of the \hmol\ molecules by a high UV radiation field from the intense star formation in the host galaxy (e.g. Hatsukade et al. 2014); observational biases against sightlines with high dust columns and against high-metallicity environments  (e.g. Ledoux et al. 2009; Covino et al.~2013); and formation of stars from atomic gas before \hmol\ has a chance to form (e.g. Michalowski et al. 2016).

An alternative approach to detecting molecular species in GRB host galaxies is through emission line spectra. This has the added advantage of avoiding
problems in interpreting line of sight measurements (e.g. the effects of ionisation and excitation by the GRB emission) and can help to place the line of sight absorption in an integrated, or in low-z cases spatially resolved (e.g. \citealt{Hatsukade}), galaxy context (\citealt{Michalowski100316D}). To date, a handful of host galaxies have been detected
in molecular line emission, in all cases this is through emission lines of carbon monoxide (CO)  (\citealt{Hatsukade}; \citealt{080517CO}; \citealt{Michalowski2016}; \citealt{Arabsalmani}; \citealt{Michalowski2018}). The use of CO as a tracer molecule for \hmol\ is a well established technique, though evidence that the CO to \hmol\ conversion factor in
GRB host sightlines is comparable to Galactic translucent clouds, is limited to a single case (\citealt{Prochaska080607h2}): the only afterglow spectrum so far with
a detection of CO absorption lines. The metallicity dependence of the CO to \hmol\ conversion factor, and other environmental effects (e.g. \citealt{Bolatto}), make the CO to \hmol\ conversion factor (and therefore a clear picture of whether GRB hosts are deficient in molecules or not) for GRB sightlines uncertain (e.g. \citealt{Arabsalmani}; \citealt{Michalowski2018}). In addition, most of the host galaxies with detected CO emission lines have a detection of only a single transition. These reasons, together with the low detection rate of CO absorption in optical afterglow spectra, makes a direct comparison between host galaxy CO emission and afterglow CO absorption difficult.

No detections of \hmol\ emission, through either pure rotational or ro-vibrational transitions, have been reported to date. 
In this paper we perform a first exploratory search for ro-vibrational \hmol\ lines in a sample of four, low redshift, long GRB host galaxies, to inform more sensitive searches with future observatories.   

\begin{table*}
 \centering
  \caption{ Observations used in this paper. $^*$: Note that the spectroscopic observations discussed
in this paper concern the brightest star forming region in this host galaxy, known as source A 
(\citealt{100316DStarling}). The magnitudes  given in this table are all integrated magnitudes 
for the whole host galaxy.  References for spectroscopy data: [1] Watson et al. (2011), [2] Wiersema (2011), 
[3] This work, [4] Starling et al. (2011). References for the host infrared magnitudes: [5] Prochaska 
et al. (2004), [6] Hjorth et al. (2012), [7] This work, [8] Micha{\l}owski et al. (2015). References for
abundance: [9] Guseva et al. (2011), [10] Wiersema et al. (2007); [11] Stanway et al. (2015a), [12] Starling et al. (2011). 
 \label{table:obslog}}
 \begin{tabular}{llllll} 
  \hline
GRB host         &   Instrument           &  Obs date     &  Redshift  & Host IR (Vega) magnitude  & $12+\log({\rm O/H})$ \\ 
 \hline  
 031203            & VLT X-Shooter      &  17 March 2009 [1]  &  0.105    & $K' = 16.54 \pm 0.02$   [5]  &  8.20 [9] \\
 060218            & VLT ISAAC           &  17 July + 10 Sep. 2008  [2]                   &  0.033  &  $K_s = 17.94 \pm 0.09$ [6]    &  7.54 [10]  \\
 080517            & WHT LIRIS           & 3/4 March 2015 [3] &  0.089  &  $K_s = 15.51 \pm 0.06$ [7]   &  $\sim$8.7 [11] \\
 100316D$^*$  & VLT X-Shooter       &  17 March 2010 [4]     &  0.059   &  $K_s = 15.93 \pm 0.09$ [8] & 8.23 [12]  \\
\hline
\end{tabular}
\normalsize
\end{table*}

\section{Observations}\label{sec:obs}
In this paper we use a small sample of four low-redshift GRB hosts as a pilot study. All four have $z\lesssim0.1$ (Table 1), to ensure a chance of detecting the 1-0 S(1)  transition in the usable range of near-infrared (NIR) spectrographs. Such low redshift GRBs are relatively rare (the mean redshift for {\em Swift}-discovered GRBs is $\sim1.9$, \citealt{Selsing}), and often hosts are too faint in the NIR range to have reasonable quality spectra at the H$_2$ wavelengths. 
To give a crude guess at the required flux limits, we used the findings of \cite{Pak}, who observed a sample of vigorously star forming galaxies,
and found that the ratio of the ro-vibrational 1--0 S(1) H$_2$ line luminosity ($L_{{\rm H}_2}$) and the far-infrared (FIR) continuum luminosity ($L_{\rm FIR}$) are broadly constant at
$L_{{\rm H}_2} / L_{\rm FIR} \sim 10^{-5}$ for a wide range in galaxy mass. The host of GRB\,031203 has a well determined FIR luminosity (\citealt{Herschel031203}),
which gives an expected \hmol\ flux of a few times $10^{-17}$ erg\,s$^{-1}$\,cm$^{-2}$.

For three of the hosts in our sample, the spectra, their acquisition, reduction and calibration have been described in detail in previous papers: the hosts of GRBs 060218 (\citealt{060218}; \citealt{060218IR}), 031203 \citep{031203} and 100316D (\citealt{100316DStarling}; \citealt{060218IR}; Flores et al. in prep). The host of GRB\,100316D is large, and we use the spectrum of 
the brightest star forming region in this host (a.k.a. source ``A'', \citealt{100316DStarling}).  

\begin{figure}
\centerline{\includegraphics[width=7.5cm]{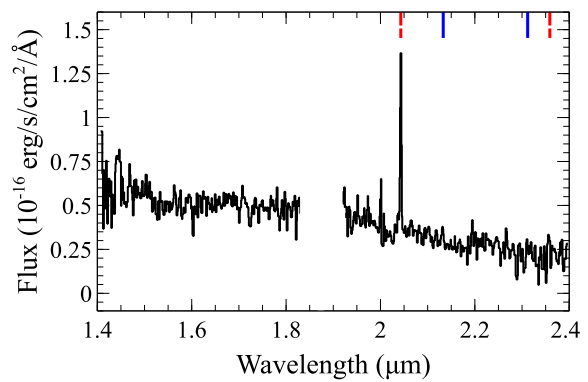}}
\caption{The LIRIS spectrum of the host galaxy of GRB\,080517. For plotting purposes the spectrum is smoothed with
a 3 pixel median filter. The wavelength is as observed. The area
worst affected by telluric absorption (i.e. with strongest residuals after telluric absorption correction) is omitted
from the plot. The dashed red lines indicate the wavelengths of 
 Paschen $\alpha$ (left) and Brackett $\gamma$ (right); the solid blue lines indicate the wavelengths of the \hmol\ 1--0 S(3) (left) and 1--0 
S(1) transitions.}
\label{fig:080517}
\end{figure}

In addition to these sources, we observed the host of nearby GRB\,080517 \citep{080517}. This last source is of particular interest because of a detection of 
an emission line of carbon monoxide (\citealt{080517CO}). 
We observed this host with the Long-slit Intermediate Resolution Infrared Spectrograph (LIRIS, \citealt{LIRIS}) on the 4.2m William Herschel Telescope,
starting at 23:31 UT on 3 March 2015. We used the low 
resolution HK grism ({\em lr\_hk}) and a 1 arcsecond wide slit, which gave a wavelength range $1.089 - 2.396$ $\mu$m. 
We obtained 4 nodding cycles, of 2 positions each,  using an exposure time of 450 s per exposure. Seeing conditions during the observations varied between 1.1 and 1.9 
arc seconds, with the latter value measured on $H$ and $K_s$ images taken directly after the science exposures.
We reduced the data using version 2.15 of the {\em lirisdr}\footnote{{\em lirisdr} is supported by J. Acosta.} package in {\sc IRAF}\footnote{{\sc IRAF} is distributed by 
National Optical Astronomy Observatories, operated by the Association of Universities for Research in Astronomy, Inc., under contract with the National Science 
Foundation.}. We observed the star SAO 013747 (an A0 photometric standard star) to aid flux calibration: correction for telluric features, and flux calibration, was done using 
the {\em SpeXtool} software package (\citealt{SpexTool}), in particular the {\em xtellcor\_general} task. To bring the spectrum onto an absolute flux scale, we used LIRIS 
imaging observations in $J,H$ and $K_s$ filters. These imaging observations were performed on the night of 5 December 2014, starting at 23:24 UT, and consisted
of two cycles of 5 dither positions with 30 seconds integration time for the $H$ band, and three cycles of 5 dither positions in the $K_s$ band. The seeing was fair
at 1.2 arc seconds FWHM. After source extraction using {\sc SExtractor} (\citealt{Bertin}), calibration onto 12 bright stars in the 2MASS survey gives the following
 magnitudes (in the 2MASS Vega system) for the host galaxy: $H = 15.78 \pm 0.05$ and $K_s = 15.51 \pm 0.06$. The response and telluric absorption calibrated
 host galaxy spectrum was corrected using these photometric values, and is shown in Figure \ref{fig:080517}. 

Generally speaking, the spectra we use in this paper were obtained as part of GRB follow-up campaigns, and therefore are a heterogeneous sample in
depth, wavelength range and resolution.

\begin{figure}
\centerline{\includegraphics[width=9cm]{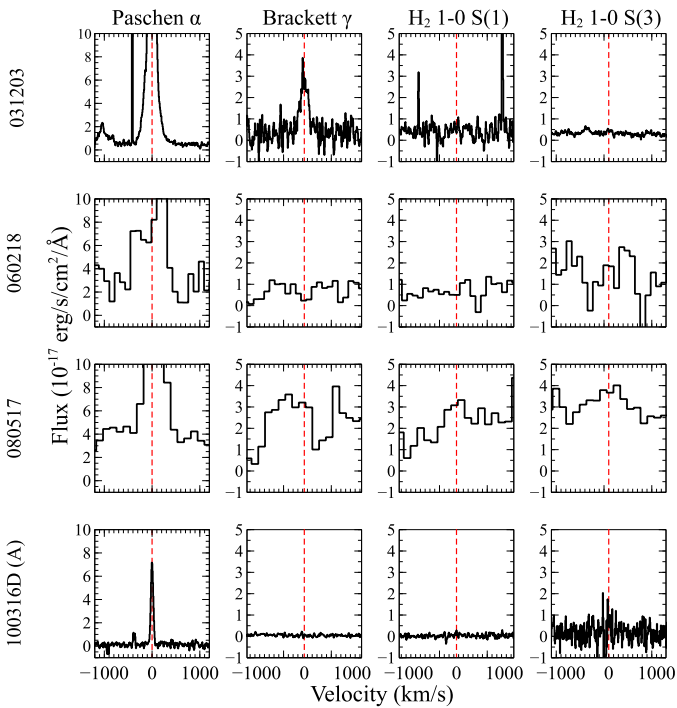}}
\caption{This plot shows postage stamp cut-outs of the spectra of the four host
galaxies in this paper. Each spectrum covers  the range $-1200$ to $+1200$ km/s around the
position of four lines of interest, which are indicated with a vertical red dashed line. 
The flux axis is identical for all lines and sources, with the exception of the Paschen\,$\alpha$
line panels. The different continuum brightness between sources is readily apparent, as well
as the varying continuum noise as lines fall in or out of regions with strong telluric absorption. A zoom-in of the top-right panel, with the only detection of a ro-vibrational \hmol\ line (1--0 S(3)) in the sample, can be seen in Figure 3.}
\label{fig:sample}
\end{figure}

\begin{figure}
\centerline{\includegraphics[width=7cm]{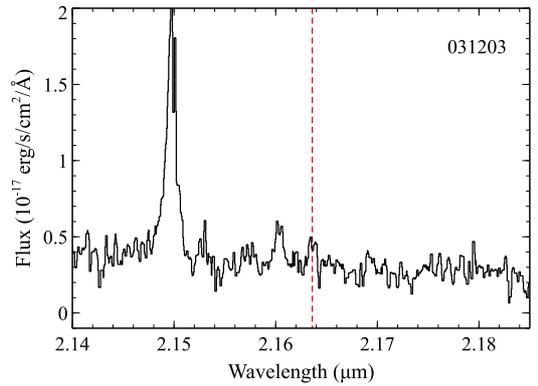}}
\caption{X-Shooter spectrum of the host galaxy of GRB\,031203, around the location
of the H$_2$ 1--0 S(3) transition, whose wavelength at the host redshift is marked with
a dashed red line; the wavelength is as observed. 
The bright emission line near 2.15\,$\mu$m is the H\,{\sc i} 8-4 (Brackett $\delta$) line, the one near 2.16\,$\mu$m is a He\,{\sc I} line.
}
\label{fig:031203detection}
\end{figure}

\section{Analysis}
For each spectrum, we used a bright emission line in the data to fix the redshift as accurately as possible: this is generally 
Paschen $\alpha$, Brackett $\gamma$ or He\,{\sc i} $\lambda$2.058 (lines close in wavelength to the wavelengths of the brightest \hmol\ lines). 
The reason for doing this, is that in some cases small velocity offsets may be present in the wavelength calibration compared to redshift values in literature.
In the mid-resolution (e.g. X-shooter) data of some hosts, in particular GRB\,031203, the emission line profiles deviate somewhat from a Gaussian profile (previously noted by \citealt{Guseva}) for the brighter lines, a common property for GRB host galaxy spectra (see e.g. \citealt{060218}). This can generally be attributed to kinematics within the host (e.g. \citealt{Arabsalmani2018velocity}). In the flux measurements (and determination of upper limits) of the \hmol\ lines in the X-shooter spectra, we make the assumption that the \hmol\ line profile is the same as that of the Paschen and/or Balmer lines. 

After the redshift was found accurately, we used a catalogue of \hmol\ ro-vibrational transition wavelengths (e.g. \citealt{BlackvanDishoeck}) to search for emission lines. In 
cases of non-detection, we computed 3$\sigma$ upper limits on the line flux, using the emission line profile properties of Paschen $\alpha$ and/or Brackett 
$\gamma$. We note that the absolute flux calibration of the spectra has a considerable uncertainty: the slit may not encompass the entire extent of the galaxy, some
data were taken during somewhat non-spectrophotometric conditions, and telluric absorption often complicates flux calibration uncertainties. 
Line flux {\em ratios} are more reliable, especially of lines close together in wavelength (and similarly affected by telluric absorption), and we therefore primarily express our limits and detections as a ratio, using
the bright Pa $\alpha$ line which is detected at high signal-to-noise in all four sources.

Figure \ref{fig:sample} shows some postage-stamp cutouts of the spectra at the positions of a selection of useful lines. It is clear from these that the signal-to-noise for a given source can vary a lot with wavelength: the different redshifts of the host galaxies shift lines nearer or further from telluric absorption and emission features (the influence of these is stronger for the low resolution spectra), or further into, or out of, wavelength regions where the spectrograph sensitivity is poorer (e.g. in X-shooter data the thermal noise in the $K$ band is high, \citealt{VernetXSH}).  Generally speaking, we limit our search to a dozen lines bluewards of $\sim2.25$ $\mu$m, the
strong Q branch lines at $\sim2.4$ $\mu$m and the 2--1 S(1) line (useful diagnostic for collisional excitation, \citealt{BlackvanDishoeck}, see section \ref{section:discussion}) are unfortunately not in reach for these sources with the spectrographs we used.

While long GRBs are (generally) accompanied by highly energetic type Ib/c supernovae (though not always detected, \citealt{FynboSNless}), the contribution of emission of such SNe to our spectra is negligible: our spectra are either taken long after the SN has faded away (Table \ref{table:obslog}), or, in the case of 100316D, cover a region away from the GRB/SN (\citealt{100316DStarling}).

All spectra contain a high significance detection of Paschen $\alpha$. For three out of four hosts we do not detect any emission lines with significance $>3\sigma$ at the positions of the \hmol\ transitions. In the Xshooter spectrum of the host of GRB\,031203, we detect a single \hmol\ transition, 1--0 S(3), with flux significance of 
$\sim4 \sigma$ (Figure \ref{fig:031203detection}), and with the expected centre wavelength and line shape. We estimate its flux at $\sim1.0 \pm 0.24 \times10^{-17}$ erg\,s$^{-1}$\,cm$^{-2}$,
with the caveats applying to the line fluxes as listed above.

The results of the flux ratio measurements for the lines that are expected to be strongest (see below) are given in Table \ref{table:fluxes}.

\section{Rotational lines in the host of GRB~031203}
The host of GRB\,031203 is one of the very few host galaxies with good quality 
(mid-)infrared spectra taken by {\em Spitzer}, using the Infrared Spectrograph (IRS) instrument (\citealt{031203}).  These spectra (shown in Figure 1 of \citealt{031203}) cover a range from $\sim 5-35 \mu$m in the rest frame, and therefore cover strong pure rotational transitions of \hmol, in particular the S(0) to S(7) transitions. Motivated by the possible detection of a ro-vibrational \hmol\ line, we reanalyze these {\em Spitzer} spectra here. 

\subsection{Model fitting and molecular mass estimates}\label{sec:pahfitresults}
Considering the diverse range of heating environments in the galaxy interstellar medium (ISM), the traditional method of fitting two or three discrete temperature molecule components to the excitation diagram is not realistic. Instead we assumed a continuous power law temperature distribution for H$_{2}$ to fit the excitation diagram and hence calculated the total H$_{2}$ gas mass in the ISM (for details refer \citealt{Togi16}). We assumed that the column density of H$_{2}$ molecules is distributed as a power law function with respect to temperature, $dN \propto T^{-n} dT$, where $dN$ is the number of molecules in the temperature range $T$ to $T+dT$. The model consists of three parameters, upper and lower temperature with power law index, denoted by $T_{\rm u}$, $T_{\rm l}$, and $n$, respectively. 

\begin{figure}
\centerline{\includegraphics[width=7cm]{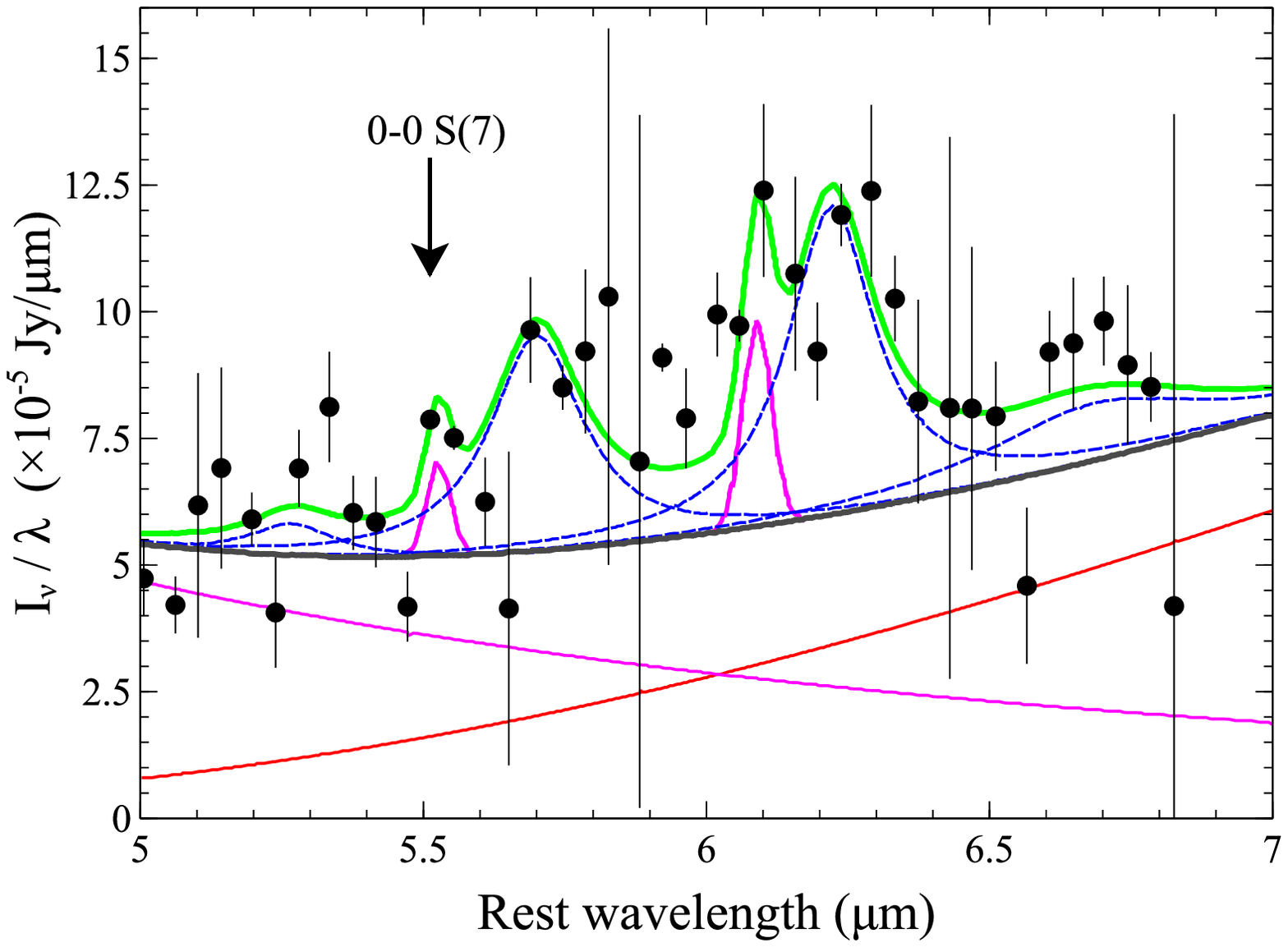}}
\caption{A small section of the {\em Spitzer} spectrum with the PAHFIT fit is shown (note that the model is fit over the full {\em Spitzer} range, see Section \ref{sec:pahfitresults}). We use the standard PAHFIT convention for the components (see Smith et al.~2007): the black-body fit to the (dust) continuum (red), the stellar continuum (magenta), the combined continuum (thick grey), the PAH features (blue), the narrow atomic and molecular features (narrow magenta components), the composite fit (green) and the datapoints (black). The location of the 0-0 S(7) transition is indicated with a vertical arrow, the other magenta peak is the 0-0 S(6) transition (see Section \ref{sec:pahfitresults} for a discussion).}
\label{fig:pahfit}
\end{figure}

We attempted to fit the mid-infrared (MIR) spectrum for the host of GRB\,031203 using {\sc PAHFIT}, a spectral decomposition tool to estimate the H$_{2}$ line fluxes \citep{Smith07}. 
No convincing H$_{2}$ line flux was detected, except a tentative detection of the 0--0 S(7) line, with a flux of 
$\sim5.3\times10^{-16}$ erg\,s$^{-1}$\,cm$^{-2}$.
A section of the spectrum and the fitted components, including the 0--0 S(7) transition, is shown in Figure \ref{fig:pahfit}. As is clear from Figure \ref{fig:pahfit}, the 0--0 S(7) covers only a few datapoints with low flux errors, and we consider its detection tentative for that reason.
The 0--0 S(6) transition (the magenta component at 6.11 $\mu$m in Fig.~\ref{fig:pahfit}) is very close to a strong polycyclic aromatic hydrocarbon (PAH) feature at 6.22 $\mu$m, and the  uncertainties in the datapoints are larger than for 0--0 S(7). We therefore don't consider the 0--0 S(6) line reliably detected, and will in the following only consider the tentative detection of the 0--0 S(7) line.

The power law index could not be determined due to non detection of any other rotational H$_{2}$ lines. The average power law index $n$ is about 4.84$\pm$0.61 in the the galaxy sample of \cite{Togi16} and $\sim$4.2 in shock regions of Stephan's Quintet \citep{Appleton17}. Using equation 9 of \citealt{Togi16} and assuming a value of $n = 4.5$, $T_{\rm l}$ = 50 K, $T_{\rm u}$= 2000 K, with the S(7) line flux estimate above, the calculated H$_{2}$ gas mass is 1.7$\times$10$^{9}$ M$_{\odot}$ in the host of GRB\,031203 (2.3$\times$10$^{9}$ M$_{\odot}$, including He and heavy element mass corrections). The variation in $n$ from 4.0--5.0 can change the gas mass in the range (0.37--7.5)$\times$10$^{9}$ M$_{\odot}$ (and (0.5--10)$\times$10$^{9}$ M$_{\odot}$ with He and heavy element correction). The upper dust mass limit estimated is 10$^{8}$ M$_{\odot}$ in this host \citep{Michalowski100316D}. Using a gas-to-dust ratio of 500, using the \cite{Remy14} scaling relation for the metallicity value of 8.2 for GRB 031203, the limit on the gas mass is 
$<5.0\times10^{10}$ M$_{\odot}$, consistent with our estimates. Our study suggests the presence of a few $\times$ 10$^{9}$ M$_{\odot}$ molecular gas in this host galaxy, but we caution that higher spectral resolution data, at higher signal-to-noise, is required to establish the veracity and accurate flux of the S(7) line candidate.
We can compare the H$_2$ mass estimate with other GRB hosts, using Figure 3 of \cite{080517CO}, and using the 
star formation rate and stellar mass from the literature. Using the values from \cite{Savaglio}, $\log(M_{*}/{\rm M}_{\odot})\sim8.8$, SFR$\sim12.7 {\rm M}_{\odot}/{\rm yr}$, we find that a H$_2$ mass around or below $10^9$ M$_{\odot}$ 
is consistent with the behaviour of other GRB hosts in this diagram. However, we note that estimates of stellar mass (and star formation rate) for this host show a large scatter, ranging from $\log(M_{*}/{\rm M}_{\odot})\sim8.4$ to $\sim9.5$ (\citealt{Guseva} and \citealt{Michalowski100316D}, respectively). 

\section{Discussion }\label{section:discussion}
The spectra of GRB host galaxies are dominated by strong emission lines from H\,{\sc ii} regions, whereas \hmol\ lines originate in colder neutral molecular clouds. The \hmol\ 
in our hosts can be excited by two main processes: through collisional processes (e.g. involving shocks from stellar winds and supernovae, rational levels are
populated by collisions of \hmol\ with neutral atoms or other \hmol\ molecules); or through fluorescence. In this latter process, the \hmol\ molecules absorb UV photons
(the Lyman-Werner bands), and then decay into ro-vibrational states. Observations of extragalactic sources have shown both processes may 
occur (e.g. \citealt{Izotov2016}; \citealt{Pak}).  The two processes can be differentiated by comparing fluxes of different  \hmol\ transitions with models (e.g. 
\citealt{BlackvanDishoeck}). In particular, UV fluorescence should give rise to brighter lines from transitions of higher $\nu$ states. 

Recently, a large NIR spectroscopic survey of blue compact dwarf galaxies and compact H\,{\sc ii} regions in nearby galaxies (\citealt{Izotov2016}; \citealt{Izotov2011}) showed that the majority of detected \hmol\ ro-vibrational line fluxes in that galaxy sample can be well explained by fluorescence. 
The similarity between the sources in \citealt{Izotov2016} and the physical properties of long GRB hosts in general (e.g. in metallicity, star formation rate, stellar age), suggests that fluorescence is likely to play a lead role in GRB host emission too. 
Comparison of the measured flux of 1--0 S(3) in the host of 031203 with the best limits on other ro-vibrational \hmol\ lines in the same source, shows broad consistency with expectations from fluorescence: we choose the same models from \cite{BlackvanDishoeck} as favoured by \cite{Izotov2016}, and use the measured \hmol\ flux ratios from 
\cite{Izotov2016} to compensate for the absence of 1--0 S(3) from the prediction tables of \cite{BlackvanDishoeck}. 
In the host of GRB\,031203, the ratio 1--0 S(1) / Br\,$\gamma$ $\lesssim0.12$  is comparable to the values found in the samples of \cite{Izotov2011}; \cite{Izotov2016}; \cite{Vanzi}.

The host galaxy of GRB\,080517 is the only one in our sample with a detection of a CO emission line (\citealt{080517CO}). It is also a somewhat unusual source for its bright NIR continuum, caused by a combination of mass, redshift and the presence of a relatively bright older stellar population in addition to the ongoing star formation, see \cite{080517}. While our WHT spectroscopy is too shallow to give useful limits on the CO/\hmol\ ratio, this source will be a key target for higher resolution IR spectrographs on bigger telescopes.

GRBs 060218 and 100316D gave rise to two of the best studied GRB associated supernovae, and are therefore important as keystones for GRB-host studies, as the supernovae may provide a direct link to stellar GRB progenitor properties. In both cases the sources are faint (Table \ref{table:obslog}; note that for 100316D the full integrated magnitude is given, not the 
one of only source A), with very weak continuum and very strong nebular lines. The host of 060218 has a very low mass and metallicity (Table \ref{table:obslog}) and
a high specific star formation, making it unlikely to host strong \hmol\ lines, as pointed out in a first search by \cite{060218IR}. The better resolution of X-shooter
would allow a more sensitive limit on the \hmol\  fluxes to be set in the future, as the redshift of this source places most lines in regions affected by strong telluric absorption.

\begin{table*}
 \centering
  \caption{ Listed are the flux ratios of the molecular line fluxes (or limits) divided by the Paschen $\alpha$ line flux. 
  Limits are $3\sigma$. A dash indicates that a line measurement was not possible due to very strong telluric absorption, 
  presence of residuals from sky emission lines, very poor signal-to-noise, or because the wavelength is not covered by the spectrum.
  $^*$: Note that the spectroscopic observations discussed
in this paper concern the brightest star forming region in this host galaxy, known as source A (\citealt{100316DStarling}).\label{table:fluxes}}
 \begin{tabular}{lllllll} 
  \hline
GRB host         & 3--1 S(1)            & 3--1 Q(1)             &1--0 S(3)                                    &1--0 S(2)                    & 1--0 S(1)                     & 1--0 S(0)                  \\
                         & (1.23 $\mu$m) & (1.31 $\mu$m)    & (1.96 $\mu$m)                         & (2.03 $\mu$m)         &  (2.12 $\mu$m)         & (2.22 $\mu$m)        \\
 \hline  
 031203            &                -        &              -             &$(2.4\pm 0.6) \times 10^{-3}$  & $<2.9\times10^{-3}$ & $<1.0\times10^{-2}$ & $<7.3\times10^{-3}$\\
 060218            &                -        &             -               &  $<0.40$                                 &       $<0.13$               &   $<0.14$                   &   $<0.18$                 \\
 080517            &                -        &             -              &  $<0.12$                                 &      $<0.13$               & $<0.29$                    &     -                            \\
 100316D$^*$  &      $<0.079$    &    $<0.14$            &  $<0.52$                                  &      $<0.14$               &  $<0.03$                    &    $<0.021$            \\
\hline
\end{tabular}
\normalsize
\end{table*}

In the coming era of highly sensitive instruments at NIR and MIR wavelengths, such as the Near Infrared Spectrograph (NIRSpec) and Mid Infrared Instrument (MIRI) instruments onboard the James Webb Space Telescope (JWST), covering wavelengths 0.6--28 $\mu$m, the ro-vibrational lines of H$_{2}$ can easily be detected (the 1--0 S(1) line will be
in the MIRI spectral range for the entire known GRB redshift distribution), and the molecular gas properties in the GRB host galaxies can be studied using multiple transitions. Telluric absorption lines strongly limit the use of ground-based low resolution spectroscopy: the change to space-based JWST data will allow access to a larger number of transitions in each spectrum. Using the flux estimate of the 1--0 S(3) line in the host of GRB\,031203, we find that NIRSpec can provide a highly significant, spatially resolved, detection of this line in this host using exposure times under an hour, as well as several other ro-vibrational transitions. The same holds for MIRI spectroscopy targetting the rotational lines in low redshift GRB hosts. 
Detecting ro-vibrational lines in GRB host galaxies that showed Lyman-Werner lines in their afterglow spectra, would require the presence of substantially larger reservoirs of warm \hmol\ than seen in the host of GRB\,031203.

\section{Conclusions}\label{sec:conclusions}
Motivated by the recent detections of CO molecule emission in GRB host galaxies, 
we searched rest-frame infrared spectra of a sample of four low redshift GRB host galaxies for signatures of H$_2$ ro-vibrational emission lines. 
A single  ro-vibrational \hmol\ emission line candidate is detected at the position of the 1--0 S(3) transition in the host of GRB\,031203. The other GRB host spectra in our sample show no significant \hmol\ line candidates, which is likely caused by signal-to-noise and resolution limitations, as well as the positions of the lines near telluric absorption features. We re-analysed low resolution {\em Spitzer} mid-infrared spectra of the host of GRB\,031203 to search for \hmol\ rotational lines. A single weak line candidate, at the position of the 0--0 S(7) transition, is seen, but the reality of this line is debatable, because of the low resolution of the {\em Spitzer} spectra. Observations with future facilities with better resolution and higher sensitivity, particularly from space, will provide the means to detect the multiple lines required for proper comparison with low-redshift galaxy samples and high redshift molecule detections in afterglow spectra.

\section*{Acknowledgements}
It is a pleasure to thank the staff at ING for their help in obtaining the LIRIS observations in this
paper, and Jose Acosta Pulido for development of (and friendly assistance with) the {\em lirisdr} software.  
We thank the anonymous referee for their constructive feedback. We acknowledge M. Micha{\l}owski and M. Arabsalmani for useful discussions. 
Based on observations collected at the European Organisation for Astronomical Research in the Southern Hemisphere 
under ESO programmes 60.A-9022(C), 381.D-0723(C) and 084.A-0260(B). The WHT and its override programme (W/2015A/11
for observations in this paper) are operated on the island of La Palma by the Isaac Newton Group in the Spanish Observatorio del Roque de los 
Muchachos of the Instituto de Astrof\'{i}sica de Canarias. AT is grateful for support from the National Science Foundation under grant no. 1616828. 
KW, NRT and RLCS acknowledge funding from STFC. The Cosmic Dawn center is funded by the DNRF.









\bsp	
\label{lastpage}
\end{document}